\documentclass[%
 a4paper,prl,twocolumn,showpacs,superscriptaddress,floatfix,nofootinbib,reprint]{revtex4-1}
\usepackage{graphicx}% Include figure files
\usepackage{dcolumn}% Align table columns on decimal point
\usepackage{bm}% bold math
\usepackage{color}
\begin{document}

%\preprint{AIP/123-QED}

\title[]{Vortical amplification of magnetic field at inward shock of supernova remnant Cassiopeia A}% Force line breaks with \\
%\thanks{Footnote to title of article.}

\author{F. Fraschetti}
 \altaffiliation{Depts. of Planetary Sciences and Astronomy, University of Arizona, Tucson, AZ, 85721, USA; Harvard/Smithsonian Center for Astrophysics, Cambridge, MA, 02138, USA}%Lines break automatically or can be forced with \\
 \email{ffrasche@lpl.arizona.edu.}

\author{S. Katsuda}
 \altaffiliation{Graduate School of Science and Engineering, Saitama University, 255 Shimo-Ohkubo, Sakura, Saitama 338-8570, Japan}

\author{T. Sato}
\altaffiliation{RIKEN Nishina Center, 2-1 Hirosawa, Wako, Saitama 351-0198, Japan}

\author{J. R. Jokipii}
\altaffiliation{Depts. of Planetary Sciences and Astronomy, University of Arizona, Tucson, AZ, 85721, USA}

\author{J. Giacalone}
\altaffiliation{Depts. of Planetary Sciences and Astronomy, University of Arizona, Tucson, AZ, 85721, USA}

%\author{...}

\date{\today}% It is always \today, today,

\begin{abstract}
We present an interpretation of the time variability of the $X$-ray flux recently reported from a multi-epoch campaign of $15$ years observations of the supernova remnant Cassiopeia A by {\it Chandra}. We show for the first time quantitatively that the $[4.2-6]$ keV non-thermal flux increase up to $50\%$ traces the growth of the magnetic field due to vortical amplification mechanism at a reflection inward shock colliding with inner overdensities. The fast synchrotron cooling as compared with shock-acceleration time scale qualitatively supports the flux decrease. 
%
%Valid PACS numbers may be entered using the \verb+\pacs{#1}+ command.
\end{abstract}

\pacs{Valid PACS appear here}% PACS, the Physics and Astronomy
%Use showkeys class option if keyword
\maketitle

Time variability of $X$-ray flux of supernova remnants enables to probe magnetohydrodynamic dynamo processes and particle acceleration at collisionless shocks \citep{Reynolds:08}. Laboratory astrophysics experiments have identified a dynamo mechanism amplifying magnetic fields at shock waves \citep{Meinecke.etal:14}; however, direct astrophysical observations have been lacking.  

A year-scale time variability in the $X$-ray filaments, or knots, of the supernova remnant Cassiopeia A was associated \citep{Uchiyama.Aharonian:08} with a fast synchrotron cooling in strong magnetic field; a decline of the $X$-ray flux between $2000$ and $2010$ was observed with {\it Chandra} in the entire remnant's western limb \cite{Patnaude.etal:11}. Recent high spatial resolution multi-epoch observations of Cassiopeia A have shown unprecedented evidence of an increase followed by a decrease of $X$-ray flux ($[4.2 - 6]$ keV band) up to $50 \%$ in six distinct regions approximately $10" \times 10"$ or $15" \times 15"$ in size located on the west side and toward the center of the remnant \citep{Sato.etal:17}; such observations cover a time period of $15$ years (from $2000$ to $2014$). The location of those regions is consistent with a high speed shock observed to move inward. Due to the young age of the remnant, such a shock is unlikely to correspond to the reverse shock, that would move outward at such evolution stage, and plausibly originated as a reflection from the collision of the forward shock with an interstellar medium molecular cloud; such a scenario was modelled in an earlier work \cite[][]{Sgro:75}. As a result of such reflection, the inward shock surface is likely to be corrugated to the scale of the molecular cloud.
 
\begin{figure}
	\includegraphics[width=0.43\textwidth]{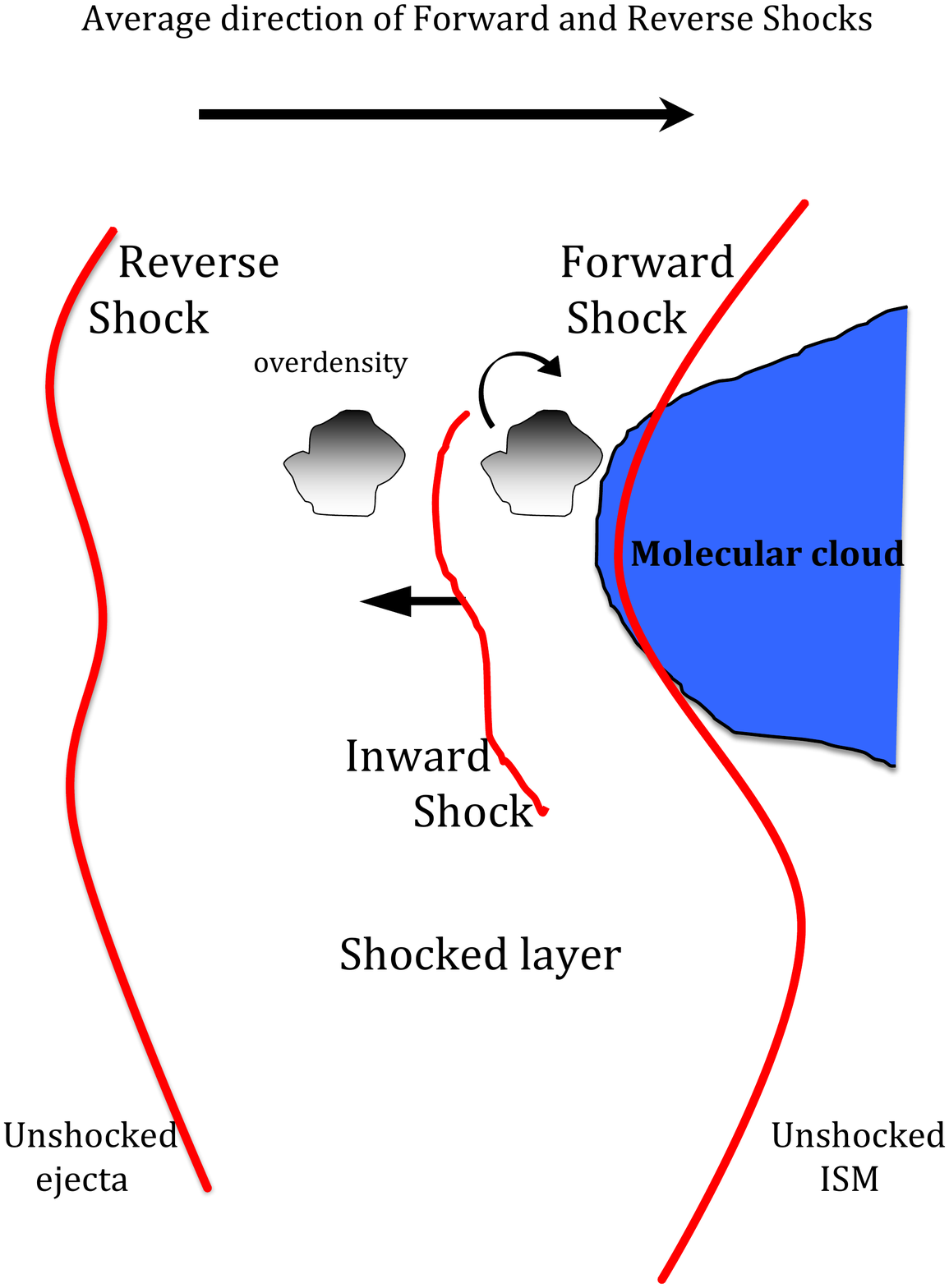}
\caption{Cartoon illustration of the proposed scenario: the inward shock recedes into the shocked layer and  crosses outward overdensity clumps thereby generating vorticity and magnetic field enhancement in the downstream fluid. The arrows indicate the direction of the shocks motion in the observer frame.}
\label{cartoon}
\end{figure}

In this paper we show for the first time that the overdensity clumps within the expanding plasma of Cassiopeia A, once crossed by the reflection inward shock, lead to the $[4.2 - 6]$ keV flux increase via magnetic field amplification through vorticity generation, as calculated in \citet{Fraschetti:13}. Such a process was first identified via two-dimensional magnetohydrodynamic (MHD) numerical simulations in \citet{Giacalone.Jokipii:07} and also investigated by several teams, including \citet{Inoue.etal:12}.   

Figure \ref{cartoon} depicts a cartoon illustration of the scenario envisaged: the corrugated inward shock travels through the shocked layer, i.e. the hot plasma region between the forward and reverse shock, and therein collides with density clumps thereby generating vorticity and amplifying the magnetic field in the downstream fluid. The  non-linear field amplification is accounted for analytically within the MHD approximation \citep{Fraschetti:13}. The  clumps might have originated in the supernova explosion itself (see, e.g., 3D simulation in \citet{Mueller.etal:17}) or from Rayleigh-Taylor instabilities triggered within the shocked layer as numerically determined in Refs. \cite{Chevalier.etal:92,Blondin.Ellison:01,Fraschetti.etal:10}. For the six regions within Cassiopeia A the speeds of the inward shocks, if propagating into the shocked layer rest frame (Fig. \ref{cartoon}), are inferred \citep{Sato.etal:17} to be in the range $5,100 - 6,800$ km$/$s.
 
The time evolution of the turbulent amplified magnetic strength $B$ is determined analytically. The induction equation for $B$ is coupled with fluid equations to determine the MHD jump conditions at two-dimensional, i.e., corrugated, shock fronts; the vorticity generated behind the shock is related to $B$ via the small-scale dynamo process \cite{Kulsrud.Anderson:92} leading to (from \citet{Fraschetti:13}, Eq. $7$ therein)
\begin{equation}
\left(\frac{B}{B_0}\right)^2 (t) = \frac{e^{2t/\tau}}{1-\alpha\tau(1-e^{2t/\tau})v_A^2 /2},
\label{exact}
\end{equation}
where $B_0$ is the upstream seed magnetic strength and 
\begin{equation}
\tau = \frac{r}{r-1} \frac{1}{C_r} \frac{R_c \ell_F}{R_c +\ell_F} 
\label{tau}
\end{equation}
is the growth time-scale determined by the shock compression $r$, the curvature radius of the ripples on the forward shock surface $R_c$ that is expected to be comparable with the size of the overdensity clumps, the  thickness of the outer clump layer where the density gradient is non-vanishing $\ell_F$ (corresponding to the Field length in the ISM) and the shock speed in the upstream frame $C_r$; $\alpha \sim 1/R_c C_r$ describes the field back-reaction to the whirling of the fluid and $v_A$ is the seed field Alfv\'en speed. As calculated in detail in \cite{Fraschetti:13}, the $B$-amplification occurs within the outer layer of thickness $\ell_F$. 

In this letter, we consider the region W3 only, according to the labelling in \cite{Sato.etal:17}, that, along with the region W1, is least affected by thermal contamination, thereby allowing a better determination of the non-thermal flux. The flux increase in the other regions (C1, C2, W1, W2, W4) moving at speeds \cite{Sato.etal:17} different from W3 is arguably produced by the vortical amplification as well, with possibly different local best-fit values of $\ell_F$ and $R_c$ to be determined in a forthcoming work.

It is reasonable to assume that the non-thermal emission arises from synchrotron radiation of a population of energetic electrons in a strong and time-varying magnetic field $B(t)$. For simplicity, we assume that over a sufficiently small interval of electron Lorentz factor $\gamma = E / m_e c^2 $ (where $m_e c^2$ is the electron rest energy), the differential energy distribution of the energetic electrons can be approximated with a simple power-law: $dN/d\gamma = N_0 (\gamma /\gamma_0)^{-p}$, where $\gamma_0$ is the injection electron Lorentz factor and the index $p$ is determined by the shock compression only, as predicted by the linear test-particle version of the diffusive shock acceleration model. It has been shown recently \citep{Fraschetti.Pohl:17a} that the baseline spectrum of the Crab nebula between radio and multi-TeV can result from a single log-parabola electron distribution, instead of commonly used multiple power-laws. 
However, within the narrow photon energy range considered here ($[4.2 - 6]$ keV), the emission spectrum from a log-parabola does not significantly depart from a power-law; thus, we choose the latter as an acceptable approximation.

{\it Flux increase} - The synchrotron power emitted by a single electron, averaged over an isotropic electron distribution, is given in the local plasma frame by $P (\gamma)= (\sigma_T c / 6 \pi) \gamma^2 B^2$, where $\sigma_T$ is the Thomson cross-section and $c$ speed of light in vacuum. The total synchrotron flux at Earth from a source at distance $d$, namely $\nu F_{\nu}$, is found by folding $P_\mathrm{syn}$ with the differential energy distribution of the electrons: $\nu F_{\nu} = {1 \over 4\pi d^2} \int d \gamma P_\mathrm{syn} N (\gamma)$. We use the monochromatic approximation, i.e., the electron power is concentrated around the characteristic synchrotron energy $\epsilon_{s} = 0.29 (3eh\gamma^2 B)/(4 \pi m_e c)$, where $e$ is the electron charge and $h$ is the Planck constant. Thus, the total flux observed at Earth, i.e., $\nu F_{\nu} \simeq {1 \over 4\pi d^2} \int d \gamma P (\gamma) N (\gamma)|_{\epsilon = \epsilon_s}$, can be recast as 
\begin{equation}
\nu F_{\nu} (\epsilon, t) = {1 \over 4\pi d^2} \frac{\sigma_T c }{12 \pi} \frac{N_0}{\cal A} \gamma_0^p \epsilon^{-\frac{p-3}{2}} B(t)^{2+\frac{p-3}{2}} 
\label{nuFnu}
\end{equation}
where ${\cal A} = [0.29 (3eh )/(4 \pi m_e c)]^{-\frac{p-3}{2}}$ is a constant. 

\begin{figure}
	\includegraphics[width=0.5\textwidth]{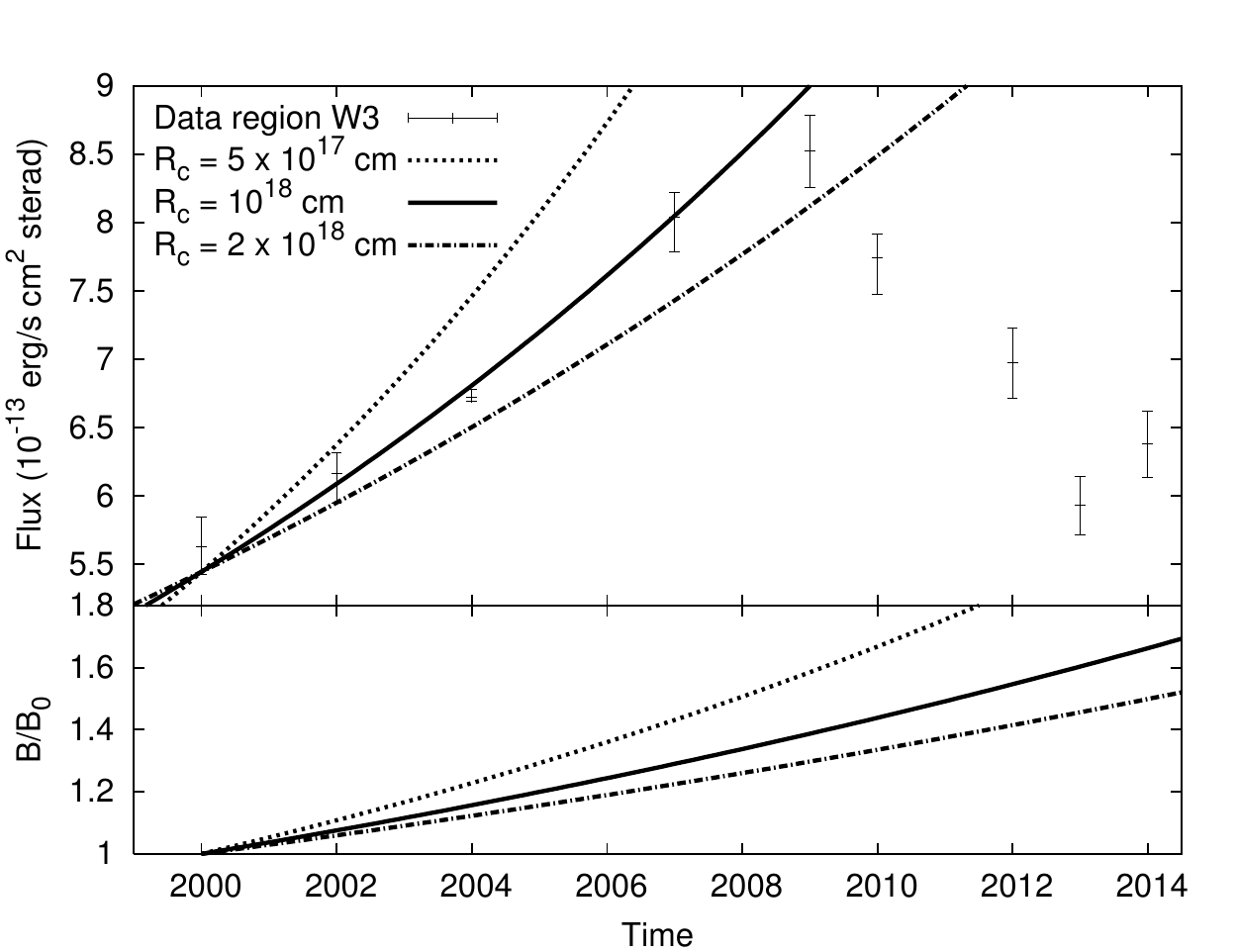}
\caption{{\it Upper panel:} Theoretical synchrotron flux at $\epsilon = 5$ keV as a function of time for distinct values of $R_c$ compared with {\it Chandra} $[4.2-6]$ keV observations; here $\ell_F = 7.03 \times 10^{17}$ cm. The best fit is represented by the solid line. {\it Lower panel:} Time-evolution of relative magnetic strength during the turbulent amplification for the three cases shown above.}
\label{flux_comp_Rc}
\end{figure}

We reproduce in Fig. \ref{flux_comp_Rc}, upper panel, the observed flux change in the range $[4.2-6]$ keV compared with the theoretical prediction (from Eq. \ref{nuFnu}) for $\epsilon = 5$ keV, $d = 3.4 $ kpc and for distinct values of $R_c$. We emphasize that $C_r$, $B$, $\epsilon$ and $r$ (hence $p$) are inferred from observations\citep{Sato.etal:17}, and are not tuned here for data-fitting purpose. The model of the flux increase depends only on two fitting parameters, $R_c$ and $\ell_F$. The best-fit values for the time-interval $[2000:2009]$ are 
\begin{equation}
R_c =  (1.00  \pm 0.16) \times 10^{18} {\rm cm} , \,  \ell_F = (7.03 \pm 0.76) \times 10^{17} {\rm cm}
\label{best-fit}
\end{equation}
The error bars on $R_c$ and $\ell_F$ are likely dominated by the uncertainty on the exponential rise. In addition, the $2009$ data-point belongs to the incipient decrease phase of the flux, hence is not accounted for by our analytic model. 

Figure \ref{flux_comp_Rc}, lower panel, depicts the amplification of the magnetic strength during the same time interval and for the same parameters as the upper panel. We show that the growth of $B$ extends beyond the flux maximum. However, this does not need to be the case and the growth of $B$ might be hampered in the deep downstream medium. We do not consider in this short letter such a possibility.  

The {\it Chandra} observations of the $X$-ray flux increase \citep{Sato.etal:17} are consistent with a scenario of an inward shock travelling into a hot plasma with electron temperature $\sim 2$ keV(see also \cite{Hwang.Laming:12}), in equilibrium with ions: low ion temperature is favoured by the narrow lines in some bright knots within the remnant \cite{Lazendic.etal:06}, despite not having been measured directly at the inward shock yet. For the W3 inward shock $C_r = 6,500$ km$/$s in the shocked layer frame, as derived \cite{Sato.etal:17} (table 3) from the proper motion speed $3,540 \pm 440$ km$/$s added to an expansion velocity $\sim 3,000$ km$/$s. Thus, for a mono-atomic gas, $r \sim 3.8$ leading to $p=2.07$.

The best-fit values in Eq. \ref{best-fit} are compatible with previously reported observations. The linear size of the W3 box ($\sim 10"$), capturing only the $X$-ray bright fraction of the extent of the inward shock where $B$-field is amplified ($\ell_F$), corresponds to $\sim 5 \times 10^{17}$ cm at distance $d$. Such size is in good agreement with the best-fit $\ell_F$ and is consistently smaller than the best-fit $R_c$  (Eq. \ref{best-fit}). For the ISM, $\ell_F$ is not strongly constrained by thermal equilibrium models \cite{Field:65}. However, our best-fit $\ell_F$ is consistent with expectations from thermal conduction models in multi-phase media, (see Ref \cite[][]{Begelman.McKee:90}, Eq. 5.1 and section V, b) therein). 

We note that {\it Chandra} observations set an upper limit of the seed field $B_0$ since the year $2000$: in other words the amplification, and the consequent flux increase, is likely to be have begun prior to $2000$ by the same mechanism from a smaller $B_0$, although high-resolution observations of previous epochs are not available. Our best-fit yields a growth time-scale (Eq. \ref{tau}) $\tau^* \simeq 29$ yr, whereas the {\it Chandra} observations captured only $\sim 9$ years of the flux increase (from $2000$ to $2009$). We note that $\tau^*$ is appreciably longer than the time scale variation of $X$-ray brightness in extreme cases previously reported, e.g., RX J$1713.7 - 3946$ from \citet{Uchiyama.etal:07} or in cases of very high Mach number shocks \citep{Fraschetti:13}. 

{\it Flux decrease } - As an approximate marker of the $\nu F_{\nu} ^{[4.2-6] \, {\rm keV}}$ peak, occurring in $\sim 2009$ for W3, we can use the ratio of the synchrotron cooling time scale, $t_{cool}$, to the acceleration time-scale, $t_{acc}$.  
 By using a reference value for Cassiopeia A, $ B_{0.1} = 5$, where $B_{0.1} = B  / {0.1 \, {\rm mG}}$., \citet{Sato.etal:17} find $t_{cool} \simeq 4$ years for the region W3. For an electron of Lorentz factor $\gamma$ emitting synchrotron power $P(\gamma)$ at a typical photon energy $\epsilon$ in an ambient downstream magnetic field $B$ we have 
\begin{equation}
t_{cool} (t) \simeq (55 \, {\rm yr} ) \, \epsilon_{keV}^{-1/2} B_{0.1}(t)^{-3/2}  \, ,
\label{tcool}
\end{equation} 
where $\epsilon_{keV} = \epsilon/{1 keV}$ and $B_{0.1} (t) = B (t) / {0.1 \, {\rm mG}}$. Equation \ref{tcool} is equivalent to Eq. 5 in Ref [\cite{Sato.etal:17}], where $B$ is taken as constant, with $\epsilon_{keV} =  1.9 \times 10^{-3} (E_{TeV})^2 B_{0.1}$ ($E_{TeV}$ being the electron energy in TeV and the typical value of $E_{TeV}$ is consistent with the value of $\gamma_0$ chosen here). 

At $\epsilon_{keV} = 5$, the value $t_{cool} = 4$ yrs yields at a time $t^\star = 2009$ the field $B_{0.1} (t^\star) \simeq 3.4$, in the expected range for Cassiopeia A, as discussed in Ref. \citep{Uchiyama.Aharonian:08}. The best-fit $B$-field (solid line in the lower panel in Fig. \ref{flux_comp_Rc}) shows a ratio $B (t^\star) /B_0 \simeq 1.35$ that leads to ${B_0}_{(0.1)} = 2.5$, with ${B_0}_{(0.1)} = B_0 / {0.1 \, {\rm mG}}$. Such a relatively high $B_0$ confirms that an amplification in W3 likely took place at the inward shock prior to year $2000$, possibly via vorticity generation within the shocked layer due to the corrugation of the inward shock or an alternative mechanism.
 
The acceleration time-scale $t_{acc}$ can be approximated by\citep{Parizot.etal:06}
\begin{equation} 
t_{acc}  \simeq 1.83 \frac{3 r^2}{r-1}{D_0 \over C_r^2} = (43.7 {\rm yr}) \frac{3 r^2}{r-1} \, k_0 \, \epsilon_{keV}^{1/2}  {B'}_{0.1}^{-3/2} C_{r,3}^{-2} 
\label{tacc}
\end{equation} 
where, for an isotropic upstream turbulence, $D_0$ is the diffusion coefficient at the electron cut-off energy and is along the average direction of shock motion and $k_0$, assumed \cite{Parizot.etal:06} to be equal upstream and downstream, is given by $k_0 = D_0 / D_B$, where $D_B$ is the Bohm diffusion coefficient at that energy.
The departure from unity of $k_0$, both $k_0 > 1$ ($k_0 < 1$) for shocks close to quasi-parallel (or quasi-perpendicular) topology within the acceleration region, indicates that the turbulence is not dominant over the seed field \citep{Fraschetti.Giacalone:12}. Finally, the upstream field, constant in time as the amplification occurs downstream in the model presented here for the flux increase (plasma kinetic instabilities are neglected), is given by ${B'}_{0.1} = B' / {0.1 \, {\rm mG}}$ and $C_{r,3} = C_r / (1,000$ km/s). We note that Eq. \ref{tacc} is equivalent to Eq. 6 in Ref [\cite{Sato.etal:17}] with ${B'}_{0.1} = B_{0.1} (t) = $ const. For the current speed $C_{r,3} = 6,5$, in region W3 we estimate $k_0 = 4.7$ from Eq. 3 in Ref [\cite{Sato.etal:17}]; for $ {B'}_{0.1} = 2.5$, Eq. \ref{tacc} provides at present $t_{acc} \sim 42 $ years, significantly greater than $t_{cool} \sim 4$ years. 
A cooling faster than the acceleration %\footnote{Likewise, in \cite{Sato.etal:17} typical values are $t_{cool} \sim 4 $ yr and $t_{acc} \sim 7 $ yr, with a constant B-field, and again $t_{acc} > t_{cool}$.} 
is consistent with the observed flux decrease. More sophisticated models for the diffusion coefficient (e.g., \cite{Giacalone:13}, \cite{Fraschetti.Jokipii:11} will not substantially change this qualitative argument (see next Section).

We emphasize that $t_{cool} (t)$ is time-variable unlike what is customarily assumed: it shortens as $B_{0.1} (t)$ increases (see Eq. \ref{tcool}). Here $B_{0.1} (t)$ is the downstream field, amplified after the shock crossing, as described in Eq. \ref{exact}. Thus, in the early phase of the {\it Chandra} observations ($\sim 2000$) or earlier, $t_{cool}$ was much greater: from Eq. \ref{tcool}, $\epsilon_{keV}=5$ at $B_{0.1}=1$ yields $t_{cool} =25$ yr, comparable to $\tau^*$. On the other hand, electrons need time to be accelerated up to the TeV before cooling on the $t_{cool}$ scale. As a result, an early-on cooling will not affect the electrons spectrum.  

{\it Discussion } -  We have presented the first quantitative theoretical model based on ideal MHD and small-scale dynamo downstream of shocks to explain the $X$-ray flux increase at a Cassiopeia A inward shock. We have provided an argument based on the ratio of cooling and acceleration times in support of the observed flux decrease. It is natural to inquire whether $t_{cool} $ exceeds $t_{acc}$ during the flux increase, as one could expect. By using the values observed/inferred ($r = 3.8$, $\epsilon_{keV} = 5$, $C_{r,3} = 6.5$ and $k_0 = 4.7$) and determined from our best-fit of the flux increase (${B_0}_{(0.1)} = 2.5$), the ratio of Eq. \ref{tacc} to Eq.\ref{tcool} leads to ${t_{acc}}/{t_{cool}} >1$ also prior to $2009$, at odds with expectations. This should not be surprising as current values of observables are used that can lead to a significant over-estimate of $t_{acc}$ in the flux increase phase, as discussed below. 

The use of $t_{acc} / t_{cool}$ as criterion to determine the flux variability requires further discussion. The determination of $t_{acc}$ is very sensitive to $k_0 \propto C_r^2/E_{\gamma, {\rm cut, keV}}$, where  $E_{\gamma, {\rm cut, keV}}$ is the electron cut-off energy (see Eq. 3 in Ref. [\cite{Sato.etal:17}] or Eq. 5 in Ref.[\cite{Patnaude.etal:11}]). However, the uncertainty on $E_{\gamma, {\rm cut, keV}}$ alone, dominated by the coarse spatial resolution of NuSTAR, seems insufficient to justify the large ratio ${t_{acc}}/{t_{cool}} $. \citet{Grefenstette.etal:15} reports $E_{\gamma, {\rm cut, keV}} \simeq 1.3$ keV (or $E_{\gamma, {\rm cut, keV}} \simeq 2.3$ keV) for the reverse (or forward) shock region whereas an unlikely larger value ($E_{\gamma, {\rm cut, keV}} \simeq 10$ keV), with other parameters unchanged, would be required to satisfy the condition ${t_{acc}}/{t_{cool}} <1$. 

Speculative arguments to justify ${t_{acc}}/{t_{cool}} > 1$ during the flux increase involve also a temporal variation of the fitting parameters that have been assumed constant throughout this work and in Ref. [\cite{Sato.etal:17}]. For instance, the shock speed $C_r$ in the acceleration region likely decreased during the collision with the overdensity. At constant $k_0$, the relation $k_0 \propto C_r^2/E_{\gamma, {\rm cut, keV}}$ implies that a greater $C_r$ would reduce $t_{acc}$ (Eq. \ref{tacc}), with $t_{cool}$ unchanged, only by requiring a greater $E_{\gamma, {\rm cut, keV}}$. On the other hand, $k_0$ might vary in time due to the growth of the magnetic turbulence and approach unity from above (below) for a quasi-parallel (quasi-perpendicular) magnetic topology\citep{Fraschetti.Giacalone:12}. So far we have considered values of $k_0 > 1$. However, an intrinsic value $k_0 < 1$ or $k_0 \ll 1$ in a region magnetically connected with the {\it Chandra} non-thermal  bright inward shock could also lead to a production of TeV electrons that would migrate a short distance before cooling by synchrotron radiation; such a $k_0 \ll 1$ would also enable a $t_{acc} < t_{cool}$. At present we cannot disprove either magnetic topology. Thus, the observational uncertainties cannot rule out ${t_{acc}}/{t_{cool}} < 1$ during the flux increase.

The flux decrease could have a different origin: the damping of the turbulent $B$-field downstream of the inward shock, e.g., due to non-linear wave interactions\citep{Pohl.etal:05}, instead of energy losses of radiating electrons. The resulting flux decrease is expected to be achromatic from the radio to the hard $X$-ray spectrum. We will investigate such effect in a forthcoming work. Another possible origin of the flux decrease is that the overdensity begins to disrupt at the $\nu F_{\nu} ^{[4.2-6] \, {\rm keV}}$ peak, as $\tau^* \sim \ell_F/C_r$ is not negligible as compared to the shock crossing time $\sim R_c/C_r$. This effect will be accounted for in future numerical simulations.

We note that two alternative scenarios for the inward shock location are suggested by \citet{Sato.etal:17} to explain the non-thermal emission, i.e., propagation through the unshocked ejecta or through the shocked layer with ion  temperature $\sim 46$ keV, out of equilibrium with electrons. In both scenarios Eq. \ref{exact} would result in ${t_{acc}}/{t_{cool}} <1$ throughout, at odds with the flux decrease. Moreover, in the latter scenario, the best-fit parameters are consistent with those in the case of ion/electron equilibrium at $\sim 2$ keV: $R_c =  (1.21  \pm 0.19) \times 10^{18} {\rm cm} , \,  \ell_F = (7.30 \pm 0.72) \times 10^{17} {\rm cm}$. However, the smaller compression ($r \sim 2.1$) would steepen the electron spectrum ($p \sim 4$) making it unlikely to accelerate up to $\sim 10$ TeV to produce the observed $X$-rays. The former scenario would require the presence of a second molecular cloud in the unshocked ejecta, beside the outer cloud that generates the inward shock. The $^{12} CO$ maps from the Heinrich Hertz Submillimeter Telescope \citep{Sato.etal:17} with velocity $\simeq -40$ km$/$s, reveal molecular clouds overlapping, in projection, to the innermost parts of the remnant, within the putative position of the reverse shock. 
In this case the relative speed of the inward shock colliding with the second cloud in region W3 would be $C_r \simeq 3,500$ km/s (see Ref [\cite{Sato.etal:17}], table 3). A value $\ell_F \simeq 3 \times 10^{17}$ cm, with unchanged $R_c$, would still lead to $\tau^* \simeq \ell_F / C_r \simeq 29$ yrs. However, observations do not allow to clearly single out such a molecular cloud; thus, we do not consider such a scenario. 
  
{\it Conclusion -} We have shown that the $X$-ray flux increase between years $2000$ and $2009$ in a small region in the west limb of Cassiopeia A can trace the enhancement of the magnetic field due to vortical amplification as formerly proposed. The scaling of the saturation value $B/B_0$ with the Alfv\'en Mach number in the upstream fluid $M_A = C_r/v_A$ is simply given\citep{Fraschetti:13} by $B/B_0 \sim M_A$. The low speed of the inward shock in the hot ejecta and the high value of $B_0$ jointly lead to a relatively low $M_A$, i.e., $M_A \simeq 3.8$ for a reasonable value of thermal proton density $n$ ($n = 0.1$ cm$^{-3}$ leads to $v_A = 1,700$ km/s); thus, only a modest field amplification can be observed. Values of the diffusion coefficient departing from the Bohm limit, that indicate relatively weak turbulence around the shock, are also consistent with the inferred modest field  amplification. We have provided a qualitative argument but not a firm theoretical model for the flux decrease between years $2009$ and $2014$. Further analysis is warranted and high-resolution follow-up monitoring of the region W3 is encouraged. This work demonstrates for the first time that the unfolding of a dynamo process formerly theoretically identified can be not only investigated in laboratory plasma astrophysics but also observed in astrophysical systems.

\begin{acknowledgments}

We thank the referees for useful feedback. FF acknowledges very helpful discussions with J. Raymond. The work of FF was supported, in part, by NASA under Grant NNX15AJ71G. This work was partly supported by the Japan Society for the Promotion of Science KAKENHI grant numbers 16K17673 (SK), and partly by Leading Initiative for Excellent Young Researchers, MEXT, Japan. 
\end{acknowledgments}

%\appendix

%\section{Appendixes}

\def \apss{{\it Astrophys.\ Sp.\ Sci.}}
\def \aj{{\it The Astronomical Journal}}
\def \apj{{\it The Astrophysical Journal}}
\def \apjl{{\it The Astrophysical Journal Letters}}
\def \apjs{{\it The Astrophysical Journal Supplement Series}}
\def \araa{{\it Annual Review Astronomy \& Astrophysics}}
\def \prc{{\it Physical\ Review\ C}}
\def \aap{{\it Astronomy \& Astrophysics}}
\def \aaps{{\it Astronomy \& Astrophysics Supplement Series}}
\def \gca{{\it Geochim. Cosmochim.\ Act.}}
\def \grl{{\it Geophysical Research Letters}}
\def \jgr{{\it Journal of Geophysical Research}}
\def \mnras{{\it Monthly Notices of the Royal Astronomical Society}}
\def \nat{{\it Nature}}
\def \physscr{{\it Physica\ Scripta}}
\def \pre{{\it Physical\ Review\ E}}
\def \physrep{{\it Physical\ Report}}
\def \planss{{\it Planetary and Space Science}}
\def \pasp{{\it Publ.\ Astron.\ Soc.\ Pac.}}
\def \pasj{{\it Publications of the Astronomical Society of Japan}}
\def \solphys{{\it Solar\ Physics}}
\def \ssr{{\it Space\ Science\ Reviews}}

%\nocite{*}
\bibliographystyle{apsrmp4-1}
\bibliography{ms}% Produces the bibliography via BibTeX.

\end{document}